\input{aipcheck}

\documentclass[
    ,final            % use final for the camera ready runs
  ]
  {aipproc}

\layoutstyle{6x9}
\begin{document}

\title{Transverse spin effects in proton-proton scattering and $Q \bar Q$ production}

\author{S.V.~Goloskokov}{
  address={Bogoliubov Laboratory of Theoretical Physics,
 Joint Institute
for Nuclear Research, Dubna 141980, Moscow region, Russia} }

\begin{abstract}
We discuss transverse spin effects caused by the spin-flip part of
the Pomeron coupling with the proton. The predicted spin
asymmetries in proton-proton scattering and QQ production in
proton-proton and lepton-proton reactions are not small and can be
studied in future polarized experiments.
\end{abstract}

\maketitle

%%%%%%%%%%%%%%%%%%%%%%%%%%%%%%%%%%%%%%%%%%%%
%% MAINMATTER
%%%%%%%%%%%%%%%%%%%%%%%%%%%%%%%%%%%%%%%%%%%%

In this report, we discuss what future facilities can be used to
ascertain the existence of the spin spin-flip part of the Pomeron
coupling. The spin structure of the Pomeron was analysed by
different authors (see \cite{gol_mod,gol_kr} and reference
therein). Its manifestation can be investigated in the elastic pp
scattering at low $t$ \cite{kopel}. We shall analyze here the
single spin asymmetries in the elastic $pp$ scattering near the
diffraction minimum and of the $Q \bar Q$ production in the $pp$
reaction. The double spin asymmetries of $Q \bar Q$ production in
lepton-proton reactions for a longitudinally polarized lepton and
a transversely polarized proton will be studied too. It will be
shown that these asymmetries are sensitive to the spin-flip part
of the Pomeron coupling and predicted asymmetries are not small,
about 10\%. They can be used to obtain information about the spin
structure of the Pomeron coupling. To study spin effects in
diffractive reactions, we use the two- gluon exchange model ,
which is directly connected with the Pomeron. On the other hand,
diffractive processes can be expressed in terms of the generalized
or skewed parton distribution (GPD) in the nucleon $F_\zeta(x)$,
$K_\zeta(x)$ \cite{rad-j}. The connection of the two-gluon model
with GPD will be shown.

The two-gluon coupling with the proton can be parametrized in the
form  \cite{golostr}
\begin{eqnarray}\label{ver}
V_{pgg}^{\alpha\beta}(p,t,x_P,l_\perp)&=& B(t,x_P,l_\perp)
(\gamma^{\alpha} p^{\beta} + \gamma^{\beta} p^{\alpha}) \nonumber\\
&+&\frac{i K(t,x_P,l_\perp)}{2 m} (p^{\alpha} \sigma ^{\beta
\gamma} r_{\gamma} +p^{\beta} \sigma ^{\alpha \gamma}
r_{\gamma})+...  .
\end{eqnarray}
Here $m$ is the proton mass. In the matrix structure (\ref{ver})
we wrote only the terms with the maximal powers of a large proton
momentum $p$ which are symmetric in the gluon indices
$\alpha,\beta$. The structure proportional to $B(t,...)$
determines the spin-non-flip contribution. The term $\propto
K(t,...)$ leads to the transverse spin-flip at the vertex.

The spin-dependent cross section of diffractive processes are
expressed in terms of the soft gluon coupling (1), which in the
case of hadron production is convoluted with the hard hadron
production amplitude. The spin asymmetries are expressed in terms
of the functions $B (t , ...)$ and $K (t , . . .)$ integrated over
the gluon transverse momentum $l_\perp$.

The helicity-non-flip and helicity-flip amplitudes of the
polarized proton off the spinless particle (a meson or unpolarized
proton) can be written in terms of the invariant functions $\tilde
B$ and $\tilde K$
\begin{equation}
\label{fnf} F_{++}(s,t)= i s [\tilde B(t)] f(t);\;\; F_{+-}(s,t)=
i s \frac{\sqrt{|t|}}{m} \tilde K(t) f(t),
\end{equation}
where f(t) is determined by the Pomeron coupling with the other
hadron. The functions $\tilde B$ and $\tilde K$ are defined by the
integrated over $l_\perp$ structures from (\ref{ver}).

There are some models that provide  spin-flip effects which do not
vanish at high energies. In the model \cite{gol_mod}, the
amplitudes $K$ and $B$ have a phase shift caused by the soft
Pomeron rescattering effect.  The  vector diquarks in the diquark
model \cite{gol_kr} generate the  $K$ amplitude  which is out of
phase with the Pomeron contribution to the amplitude $B$. The
value $|\tilde K|/|\tilde B| \sim 0.1$ found in
\cite{gol_mod,gol_kr} will be used in our estimations of the
asymmetry in diffractive hadron production.

The single spin asymmetry is determined by
\begin{equation}\label{an}
A_N \sim 2 \frac{\sqrt{|t|}}{m} \frac{\mbox{Im}(B K^*)}{|B|^2}.
\end{equation}
The models \cite{gol_mod,gol_kr} describe the experimental data on
single spin transverse asymmetry $A_N$ \cite{krish} quite well.
Thus, the weak energy dependence of spin asymmetries in exclusive
reactions is now not in contradiction with the experiment
\cite{gol_mod,akch}.

\begin{figure}[h]
  \includegraphics[height=.28\textheight]{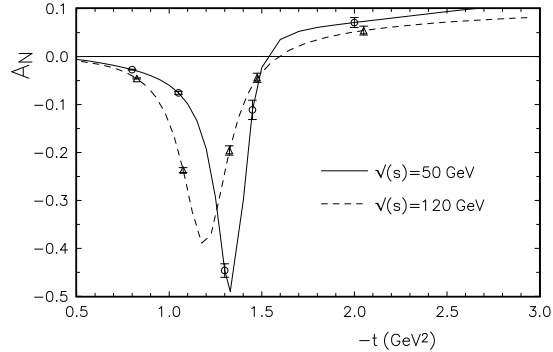}
  \caption{Predictions of the model
\cite{gol_mod} for single-spin transverse asymmetry of the $pp$
scattering at RHIC energies
  \cite{akch}. Error bar indicates expected
 statistical errors for the PP2PP experiment at RHIC.}
\end{figure}

In the diffraction minimum the imaginary part of the amplitude $B$
is equal to zero and the asymmetry is determined by the product
$(\mbox{Re}B\, \mbox{Im}K)$. Thus, the asymmetry near the
diffraction minimum is sensitive to the imaginary part of the
spin--flip $K$, amplitude and study of  the $A_N$ asymmetry in the
PP2PP experiment at RHIC can give a direct information about the
energy dependence of the  amplitude $K$. There are model
predictions for $A_N$ in this region. The model \cite{gol_mod}
predicts a large negative value of $A_N$ asymmetry near the
diffraction minimum which weakly depends on $s$ at the RHIC energy
range \cite{akch} and $A_N$ is of about 10\% for $|t| \sim 3
\mbox{GeV}^2$ (Fig.1). Other model predictions for single-spin
asymmetry at small momentum transfer was discussed in
\cite{predazzi}.

The single spin asymmetry of diffractive $Q \bar Q$ production in
polarized pp reaction was estimated in \cite{golos96}. It was
found that  $A_N$ is determined by Eq. (\ref{an}) which is
modified by the amplitude of hard $Q \bar Q$ production. The
expected value of the single spin asymmetry in this case is about
5\% in the RHIC energy range.

Let us study now diffractive double spin asymmetries of $Q \bar Q$
production in lepton-proton reactions for a longitudinally
polarized lepton and a transversely polarized proton within the
two-gluon model. This model should describe the cross sections of
hard and light quark production  at small $x<0.1$ as well. The
spin-dependent cross section can be written in the form
\begin{equation}
\label{sigma} \frac{d^5 \sigma(\pm)}{dQ^2 dy dx_p dt dk_\perp^2}=
\left(^{(2-2 y+y^2)} _{\hspace{3mm}(2-y)}\right)
 \frac{C(x_P,Q^2) \; N(\pm)}
{\sqrt{1-4(k_\perp^2+m_q^2)/M_X^2}}.
\end{equation}
Here $C(x_P,Q^2)$ is a normalization function which is common for
the spin average and spin dependent cross section; $N(\pm)$ is
determined by a sum of graphs integrated over the gluon momenta
$l$ and $l'$. The  function $N(+)$ determines the spin-average
cross section. It has the form
\begin{equation}\label{np}
N(+)=\left(|\tilde B|^2+|t|/m^2 |\tilde K|^2 \right)
\Pi^{(+)}(t,k_\perp^2,Q^2),
\end{equation}
with
\begin{eqnarray}\label{bqq}
\tilde B \sim \int^{l_\perp^2<k_0^2}_0 \frac{d^2l_\perp
(l_\perp^2+\vec l_\perp \vec r_\perp) }
{(l_\perp^2+\lambda^2)((\vec l_\perp+\vec r_\perp)^2+\lambda^2)}
B(t,l_\perp^2,x_P,...) =  { F}^g_{x_P}(x_P,t,k_0^2)\nonumber\\
\tilde K \sim \int^{l_\perp^2<k_0^2}_0 \frac{d^2l_\perp
(l_\perp^2+\vec l_\perp \vec r_\perp) }
{(l_\perp^2+\lambda^2)((\vec l_\perp+\vec r_\perp)^2+\lambda^2)}
K(t,l_\perp^2,x_P,...) =  { K}^g_{x_P}(x_P,t,k_0^2),
\end{eqnarray}
where $k_0^2 \sim (k_\perp^2+m_q^2)/(1-\beta)$. The connection of
the two-gluon structure functions from (\ref{ver}) with GPD ${
F}^g_{x_P}(x_P,t,k_0^2)$ and ${ K}^g_{x_P}(x_P,t,k_0^2)$  is
written. This connection is general and  is shown for a vector
meson and $Q \bar Q$ production in \cite{golostr}.

The function $N(-)$ which determines the spin-dependent cross
sections looks like
\begin{eqnarray}\label{nm}
N(-)=\sqrt{\frac{|t|}{m^2}} \left(\tilde B \tilde K^*+\tilde B^*
\tilde K\right) [ \frac{(\vec Q \vec S_\perp)}{m}
\Pi^{(-)}_Q(t,k_\perp^2,Q^2) \nonumber\\
 +\frac{(\vec k_\perp \vec
S_\perp)}{m} \Pi^{(-)}_k(t,k_\perp^2,Q^2)].
\end{eqnarray}
The other form of interference between $B$ and $K$ with respect to
(\ref{an}) appears in (\ref{nm}) because we consider here the
double spin effects. The large value of asymmetry will appear for
a small phase shift between the amplitudes.

The asymmetry is approximately proportional to the ratio of
polarized and spin--average gluon distribution functions
\begin{equation}\label{cltqq}
 A_{LT}^{Q \bar Q} \sim C^{Q \bar Q} \frac{{ K}^g_\zeta(\zeta)}
 {{ F}^g_\zeta(\zeta)}
 \;\;\;\mbox{with} \;\zeta=x_P\;\;\mbox{and}\; |\tilde K|/|\tilde B| \sim 0.1
\end{equation}

The spin-dependent contribution to the asymmetry which is
proportional to $\vec k_\perp \vec S_\perp$ in (\ref{nm})  will be
analyzed for the case when the transverse jet momentum $\vec
k_\perp$ is parallel to the target polarization $\vec S_\perp$.
The asymmetry is maximal in this case. To observe this
contribution to asymmetry, it is necessary to distinguish
experimentally the quark and antiquark jets.  If we do not
separate events with $\vec k_\perp$ for the quark jet, e.g., the
resulting asymmetry will be equal to zero because the transverse
momentum of the quark and antiquark are equal and opposite in
sign.
\\[8mm]
\phantom{.}
%%%%%%%%%%%%%%%%%%%%%%%%%%%%%%%5
\begin{minipage}{7.2cm}
\includegraphics[height=.24\textheight]{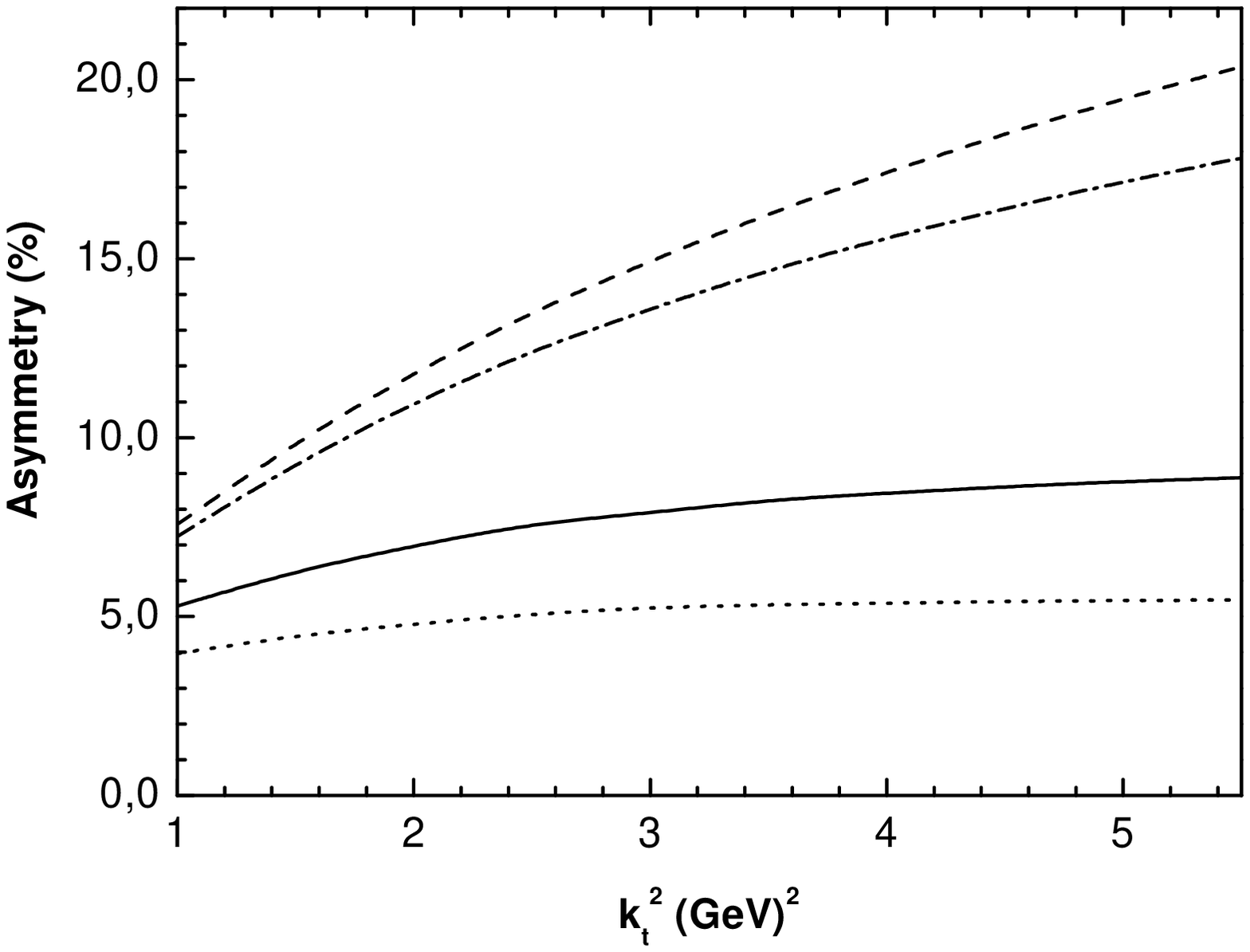}
\end{minipage}
\begin{minipage}{0.0cm}
\phantom{aa}
\end{minipage}
\begin{minipage}{7.2cm}
\includegraphics[height=.24\textheight]{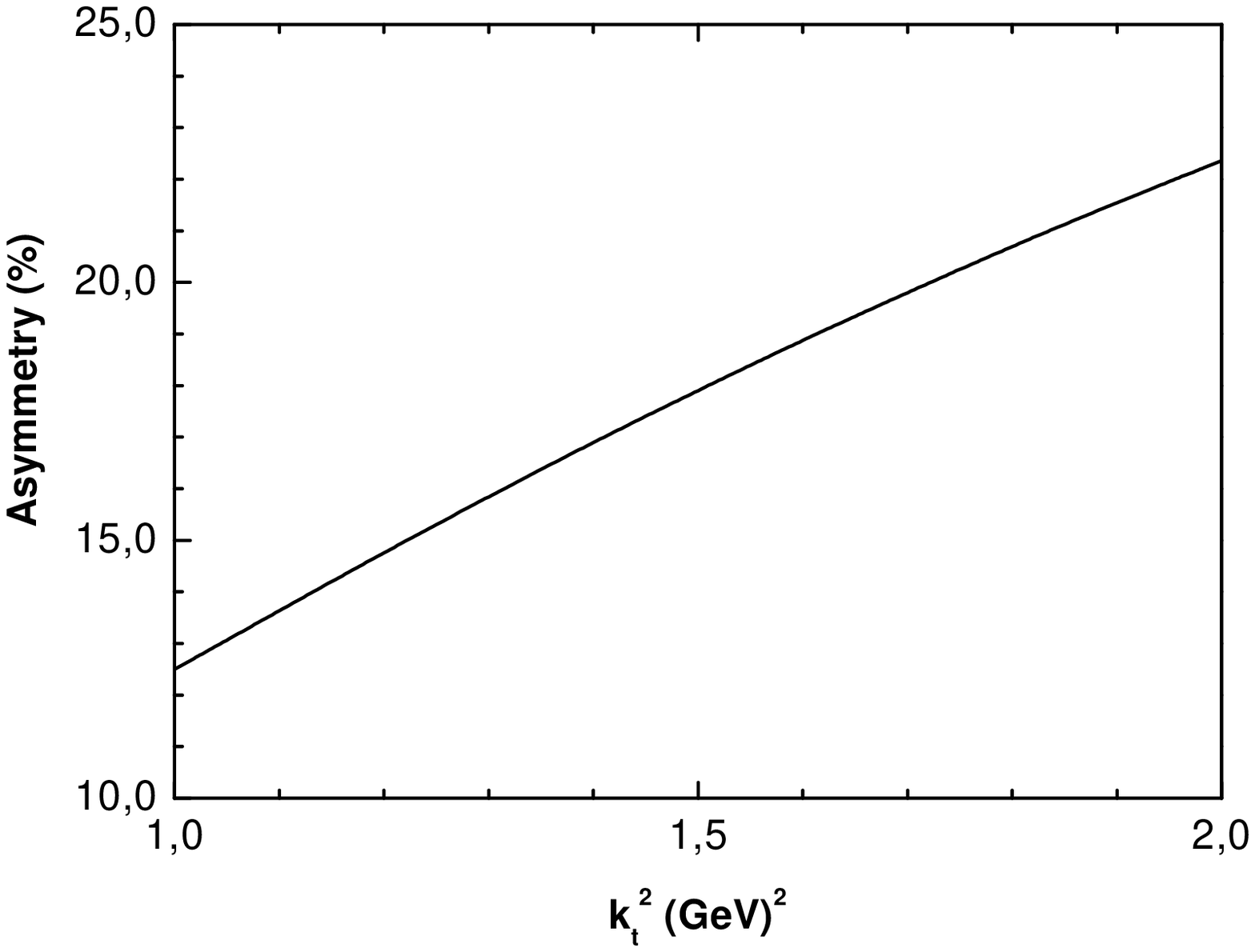}
\end{minipage}
\\[.3cm]
%\bigskip
\noindent
\begin{minipage}{7.1cm}
{\small{\bf FIGURE 2.}~ {The $A^k_{lT}$ asymmetry in diffractive
heavy $Q \bar Q$ production at $\sqrt{s}=20 \mbox{GeV}$ for
$x_P=0.1$, $y=0.5$, $|t|=0.3 \mbox{GeV}^2$: dotted line-for
$Q^2=0.5 \mbox{GeV}^2$; solid line-for $Q^2=1 \mbox{GeV}^2$;
dot-dashed line-for $Q^2=5 \mbox{GeV}^2$; dashed line-for $Q^2=10
\mbox{GeV}^2$.}}
\end{minipage}
\begin{minipage}{.15cm}
\phantom{aa}
\end{minipage}
\begin{minipage}{7.1cm}
\vspace{-1.4cm} {\small{\bf FIGURE 3.}~ {The $A^k_{lT}$ asymmetry
in diffractive light $Q \bar Q$ production for
$Q^2=5\mbox{GeV}^2$, $x_P=0.1$, $y=0.5$, $|t|=0.3(GeV)^2$ at
$\sqrt{s}=7 \mbox{GeV}$.}}
\end{minipage}
%%%%%%%%%%%%%%%%%%%%%%%%%%%%====
\\[.4cm]

The predicted asymmetry for heavy $c \bar c$ production at the
COMPASS energies is shown in Fig.2. The asymmetry for light quark
production is  approximately of the same order of magnitude. The
expected $A_{lT}$ asymmetry for light quark production at HERMES
is shown in Fig.3. The function $C_k^{Q \bar Q}$ in (\ref{cltqq})
is quite large, about 1.5 at the HERMES energy for
$k_\perp^2=1.3\mbox{GeV}^2$, $Q^2=5\mbox{GeV}^2$, $x_P=0.1$,
$y=0.5$, and $|t|=0.3 \mbox{GeV}^2$. The spin--dependent cross
section  vanishes for $Q^2 \to 0$, while the spin--average cross
section is constant in this limit. As a result, the asymmetry can
be estimated as $A_{lT} \propto Q^2/(Q^2+Q^2_0)$ with $Q^2_0 \sim
1 \mbox{GeV}^2$. This shows a possibility of studying the
polarized gluon distribution ${ K}^g_\zeta(x)$ in the COMPASS and
HERMES experiment at $Q^2 \geq 0.5\mbox{GeV}^2$.

The contribution to asymmetry $\propto \vec Q \vec S_\perp$ in
(\ref{nm}) is simpler to study experimentally. The expected
asymmetry in this case is not small too, about 5\% or a little bit
smaller. The predicted $C_Q^{Q \bar Q}$ in (\ref{cltqq}) in this
case is about 0.3. In contrast to the $A^k_{lT}$ term, the
$A^Q_{lT}$ asymmetry has a strong mass dependence \cite{golostr}.

Similar analyses have been carried out for diffractive $J/\Psi$
production which is described by the same two-gluon exchange.
Unfortunately, the double spin $A_{lT}$ asymmetry in this case is
proportional to $x_P$ which is fixed here by $x_P \sim
(m_V^2+Q^2+|t|)/(s y)$. As a result the expected asymmetry is
small and $C_g(J/\Psi) \sim 0.007$ for vector meson production in
the HERMES energy range. It is difficult to expect experimental
study of such small asymmetry.

I this report we have studied the transverse asymmetries caused by
the spin structure of the Pomeron coupling with the proton.
Similar to the low energy experiments \cite{krish} we predict the
large negative asymmetry in the vicinity of the diffractive
minimum in elastic $pp$ scattering at the RHIC energies. If the
weak energy dependence of the asymmetry in this region is found in
the PP2PP experiment at the RHIC, it will give a definite
indication of the spin-dependent Pomeron coupling. Another
possibility is to study the single-spin asymmetry of diffractive
$Q \bar Q$ production at RHIC which is predicted to be about 5\%.

The diffractive hadron leptoproduction for a longitudinally
polarized lepton and a transversely polarized proton at high
energies has been studied. The $A_{lT}$ asymmetry is found to be
proportional to the ratio of structure functions $A_{lT}=C {
K^g}/{ F^g}$. If the Pomeron has a spin structure, the ratio
$K^g/F^g$ should have a weak $x$ dependence at low $x <0.1$. The
$A_{lT}$ asymmetry can be used to get information on the
transverse distribution ${ K}^g_{x_P}(x_P,t)$ from experiment if
the function $C$ in (\ref{cltqq}) is not small. We predict that
 $C_k^{Q \bar Q} \sim 1$  for the term $\propto \vec k_\perp \vec
S_\perp$ and $C_Q^{Q \bar Q} \sim 0.3$ for the contribution
$\propto \vec Q \vec S_\perp$ in the asymmetry (\ref{nm}). We can
see that the expected values of $C$  are quite large and such
asymmetries might be excellent objects to study transverse spin
effects in the proton-- gluon coupling.

The results presented here should be applicable to the reactions
with heavy quarks. For processes with light quarks, our
predictions can be used in the small $x$ region ($x \le 0.1$ e.g.)
where the contribution of quark GPD is expected to be small. In
the case of light quark production   the polarized $u$ and $d$
quark GPD might be studied together with the gluon distribution in
the region of not small $x\ge 0.1$. Such experiments can be
conducted at the  HERMES and COMPASS spectrometers for a
transversely polarized target and future eRHIC. We conclude that
important information on the spin--dependent GPD ${ K}_\zeta(x)$
at small $x$ can be obtained from the asymmetries in diffractive
$pp$ and $lp$ reactions for longitudinally polarized lepton and
transversely polarized hadron targets.

The author is grateful to the Organizing Committee of SPIN2002 for
the local financial support. These report was supported in part by
the Russian Foundation for Basic Research, Grants 00-02-16696 and
02-02-27409.

\end{document}